\documentclass[]{spie}  

\usepackage{amsmath,amsfonts,amssymb}
\usepackage{graphicx}
\usepackage[colorlinks=true, allcolors=blue]{hyperref}

\newcommand{\rot}{\operatorname{rot}}

\title{The Infrared Imaging Spectrograph (IRIS) for TMT: motion planning with collision avoidance for the on-instrument wavefront sensors}

\author[a]{Edward L. Chapin$^*$}
\author[a]{Jennifer Dunn}
\author[b]{Jason Weiss}
\author[c]{Kim Gillies}
\author[d]{Yutaka Hayano}
\author[b]{Chris Johnson}
\author[b]{James Larkin}
\author[e]{Anna Moore}
\author[e]{Reed L. Riddle}
\author[b]{Ji Man Sohn}
\author[e]{Roger Smith}
\author[d]{Ryuji Suzuki}
\author[f]{Gregory Walth}
\author[f]{Shelley Wright}

\affil[a]{National Research Council Herzberg, 5071 W Saanich Rd, Victoria, V9E 2E7, Canada}
\affil[b]{Department of Physics and Astronomy, Univ. of California, Los Angeles, CA 90095-1547, USA}
\affil[c]{Thirty Meter Telescope International Observatory, 100 W Walnut St, \#300, Pasadena, CA 91124, USA}
\affil[d]{National Astronomical Observatory of Japan, 2-21-1 Osawa, Mitaka, Tokyo, 181-8588, Japan}
\affil[e]{Caltech Optical Observatories,1200 E California Blvd., Pasadena, CA 91125, USA}
\affil[f]{Center for Astrophysics and Space Sciences, Univ. of California, San Diego, La Jolla, CA 92093, USA}

\authorinfo{$^*$ed.chapin@nrc-cnrc.gc.ca, phone +1 250 363 6971}

\begin{document}
\maketitle

\begin{abstract}
The InfraRed Imaging Spectrograph (IRIS) will be a first-light client instrument for the Narrow Field Infrared Adaptive Optics System (NFIRAOS) on the Thirty Meter Telescope. IRIS includes three configurable tip/tilt (TT) or tip/tilt/focus (TTF) On-Instrument Wavefront Sensors (OIWFS). These sensors are positioned over natural guide star (NGS) asterisms using movable polar-coordinate pick-off arms (POA) that patrol an approximately 2-arcminute circular field-of-view (FOV). The POAs are capable of colliding with one another, so an algorithm for coordinated motion that avoids contact is required. We have adopted an approach in which arm motion is evaluated using the gradient descent of a scalar potential field that includes an attractive component towards the goal configuration (locations of target stars), and repulsive components to avoid obstacles (proximity to adjacent arms). The resulting vector field is further modified by adding a component transverse to the repulsive gradient to avoid problematic local minima in the potential. We present path planning simulations using this computationally inexpensive technique, which exhibit smooth and efficient trajectories.
\end{abstract}


\section{INTRODUCTION}
\label{sec:intro}  

\begin{figure}[ht]
\begin{center}
\includegraphics[width=0.6\linewidth]{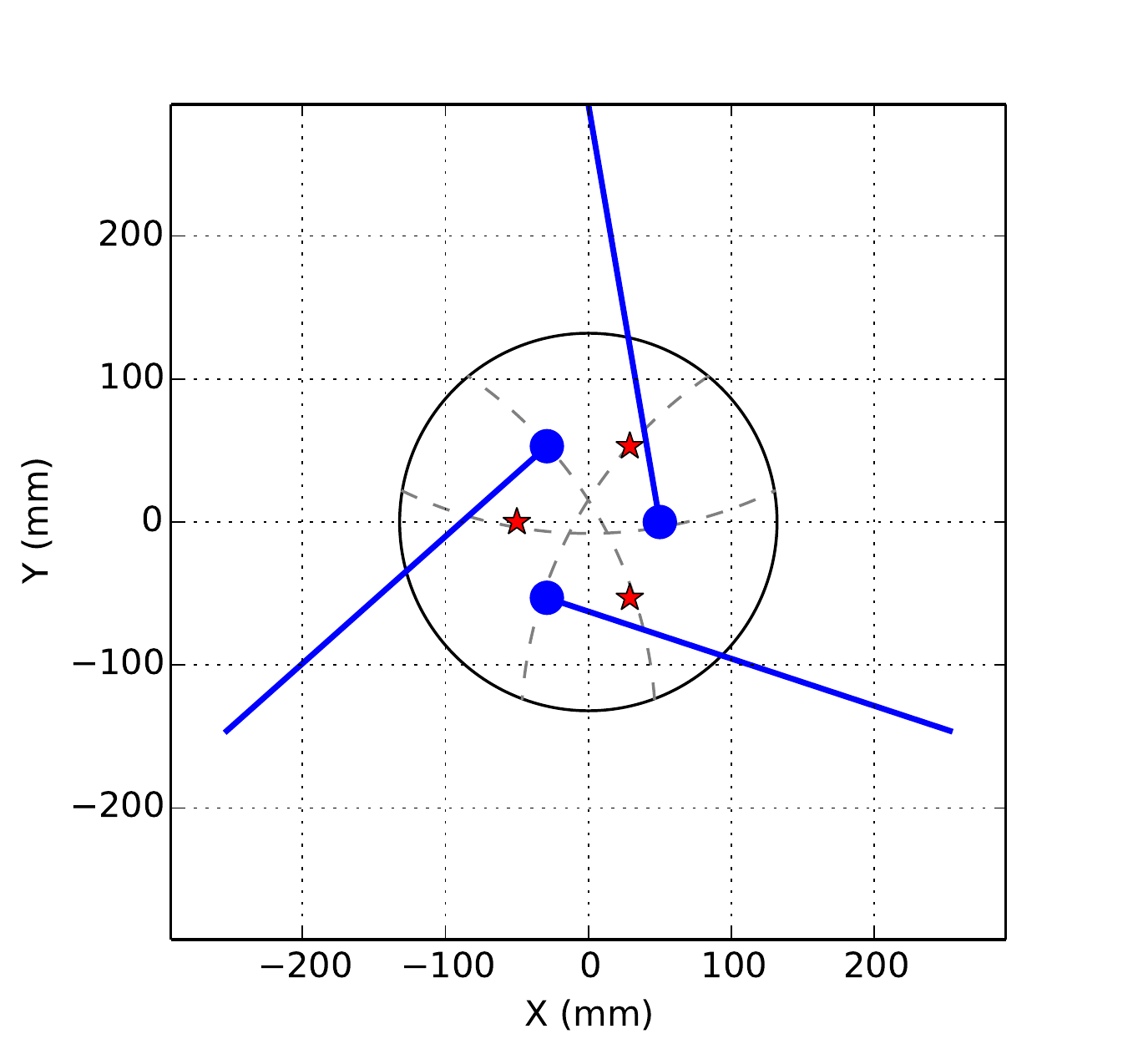}
\end{center}
\caption{\label{fig:fov}
Diagram depicting the probes at their maximum extensions (blue lines with circular heads), though rotated slightly from their home positions, a target asterism (red stars), and the probe patrol areas (bounded by their maximum extensions, shown by dashed grey lines, and the edge of the 2-arcminute FOV, indicated by the large black circle). Both 2- and 3-probe collisions are possible where their patrol areas intersect.}
\end{figure}

The InfraRed Imaging Spectrograph (IRIS)\cite{larkin2016} will be one of the first-light instruments for the Thirty Meter Telescope (TMT). Adaptive optics (AO) corrections for IRIS will be provided by the facility Narrow Field Infrared Adaptive Optics System (NFIRAOS)\cite{herriot2014}, across the full $\sim$50\,arcsec field-of-view (FOV) of the imager. In order to achieve these corrections, low-order wavefront measurements are provided to NFIRAOS using up to three configurable tip/tilt (TT) or tip/tilt/focus (TTF) On-Instrument Wavefront Sensors (OIWFS)\cite{dunn2016}. The entire OIWFS assembly is attached above the IRIS science dewar (which houses the imager and spectrograph), and is the first point of contact between IRIS and the NFIRAOS client port to which the instrument is attached. Each of the the three OIWFSs is positioned over natural guide stars (NGSs) using movable polar-coordinate (rotation and extension) pick-off-arms (POA), and together patrol an approximately 2-arcminute FOV. Being upstream of the science dewar, the POAs are generally positioned such that they avoid vignetting the smaller science FOV, although they can be configured to reach across the centre of the FOV to enable acquisition of the largest possible range of guide star asterisms (patterns of stars). This requirement on positioning flexibility results in a POA design with probes that are capable of invading each other's patrol areas; a possibility which requires mitigation as part of the positioning algorithm. Fig.~\ref{fig:fov} shows the OIWFS probe configuration, and regions in which collisions can occur (model parameters are provided in Table~\ref{tab:model}, based on an initial version of the Preliminary Design). The most likely collisions involve two probes (possible over large portions of the FOV), although three probes can also collide within a small area at the center of the field.

\begin{table}[ht]
\caption{OIWFS model parameters. Note that the minimum inter-star separation is also adopted as the minimum star-probe separation (i.e., such that a given probe does not interfere with stars being tracked by the other probes).}
\label{tab:model}
\begin{center}
\begin{tabular}{|c|c|} 
\hline
Description & Value \\
\hline
radius of patrol area              & 132\,mm \\
distance of probe origin to center & 292\,mm \\
maximum probe extension            & 300\,mm \\
radius of probe head               & 10\,mm \\
minimum inter-star separation      & 22\,mm \\
inverse plate scale                & 2.2\,mm/arcsec \\
minimum tracking speed             & 13.2\,mm/s\\
\hline
\end{tabular}
\end{center}
\end{table}

Operation of the OIWFS, from a mechanical positioning point of view, consists of two basic operations: (i)~initial configuration of the probes, which includes movement of the probes to their starting locations over guide stars; and (ii) closed-loop observations, during which corrections to the POA positions are continuously received [based on telescope motion reported by the telescope control system (TCS), or other corrections provided by NFIRAOS]. The positional corrections in (ii) are generally miniscule, as the field rotation that results from tracking sidereal objects is compensated for by the IRIS instrument rotator (gyrating the entire camera with respect to NFIRAOS, which is itself fixed to a Nasmyth platform). In this sidereal tracking case, the positional corrections will be limited to things like the nearly static offsets induced by per-OIWFS atmospheric dispersion compensators (ADC) (which are functions of the atmospheric conditions, observing geometry, guide star spectra, and the dispersive properties of the adopted ADC glasses). However, IRIS will also support non-sidereal tracking, in which the science target will remain fixed while the POAs follow their moving guide stars across the FOV. Furthermore, if a guide star moves beyond the FOV, the relevant POA will need to acquire a new guide star while the remaining probes continue tracking.

We express the POA control algorithm in terms of time-varying trajectories for each of the three probes, $\vec{q}_i(t)$, where the positions are expressed either as $\vec{q} = (r,\theta)$ in polar coordinates (relative to the origin of the probe in question), or $\vec{q} = (x,y)$ in Cartesian coordinates (absolute position in the OIWFS focal plane), and $i$ ($=1,2,3$) enumerates the probes. These trajectories connect the current, $\vec{q}_i$, and goal, $\vec{q}_{i,\mathrm{goal}}$ states. In addition to avoiding collisions, these trajectories should be: optimal, in the sense that they minimize travel time (i.e., to reduce observing overheads); smooth, to avoid velocity discontinuities which are physically impossible to follow using motor control feedback loops; and fast to calculate, so that the lag between receiving new demand positions and issuing new motor control commands is minimized (in order to support the closed-loop operational scenario). It should be noted that there are two basic strategies that span the space of possible trajectories. The first (providing an upper-bound on travel time) is a simplistic ``retract-and-rotate'' (RAR) algorithm: all of the probes can simply be retracted to a safe position at which there is no possibility of inter-probe collisions; the rotation stages may then shift to the goal angles, and then the probes can be safely extended. The other is the theoretically optimal case in which the global minimum in travel time is obtained (e.g., using a numerical optimization procedure). While the former is clearly easy to calculate (though generally resulting in less efficient trajectories), the latter is potentially quite slow due to the size of the parameter space.

Such collision avoidance problems are not new in the context of astronomical instrumentation. For example, multi-object spectrographs can have multiple robots positioning fibers within the same work space to improve observing efficiency. Following literature searches, it appears that most solutions tend to invoke simple, independently-planned trajectories for their robots, with modifications made dynamically when imminent collisions are detected\cite{goodwin2014,sugai2015}. For the OIWFS, one such solution would be to revert to RAR only when direct moves would fail.

Instead, as part of the IRIS Preliminary Design Phase, we are pursuing a widely-used robotic control strategy which makes use of an artificially constructed potential function of the state parameters. This function encodes information about the goal configuration, as well as obstacles. By following a gradient-descent trajectory through the potential, it is possible to safely move the probes towards their targets. This strategy is a good compromise between the two bounding cases mentioned above, and can be used, without modification, for both the initial configuration and closed-loop operational scenarios. We note that our approach is similar to that proposed recently to control a large array of ``$\theta-\phi$'' positioners for a fiber-fed spectrograph\cite{makarem2014a}, due to its efficiency and scalability.

\section{Motion planning using potential functions}

Our starting point is a scalar potential that includes: (i) a repulsive component, $U_\mathrm{rep}$,consisting of large positive values wherever an obstacle is located; and (ii) an attractive component, $U_\mathrm{att}$, which has its global minimum in the state space at $\vec{q}_\mathrm{goal}$. A good two-dimensional analogy is a ball rolling across a landscape: the deepest valley would be constructed at the goal location of the ball, and mountains represent the locations of obstacles to be avoided. Motion towards the goal can then be achieved by descending the gradient from the ball's starting position; assuming that the ball does not encounter any local minima along the way, it will eventually reach the goal.

We adopt typical functional forms for these two components as described, for example, in Ref.~\citenum{latombe1991}. First, the repulsive potential experienced by the $i$th probe, $U_{i,\mathrm{rep}}$, is a function of the minimum distance between it and the other probes, $D(\vec{q}_i,\vec{q}_j), i \ne j$. It is zero beyond a separation distance $Q^*$, and rapidly diverges at smaller distances:
\begin{equation}
R_{i,j}(\vec{q}_i) = \left\{
  \begin{array}{lll}
    \left( \frac{1}{D(\vec{q}_i,\vec{q}_j)} - \frac{1}{Q^*}\right) &,&
      D(\vec{q}_i,\vec{q}_j) \le Q^* \\
    0 &,& D(\vec{q}_i,\vec{q}_j) > Q^*
   \end{array}
  \right.
\end{equation}
\begin{equation}
  U_{i,\mathrm{rep}}(\vec{q}_i) = \alpha \sum_{i \ne j} R_{i,j}(\vec{q}_i)
\end{equation}
In this work we model each probe as a line segment with a circular head for the purpose of calculating $D(\vec{q}_i,\vec{q}_j)$. The parameter $\alpha$ is a tunable weight.

The attractive component as experienced by the $i$th probe, $U_{i,\mathrm{att}}(\vec{q}_i)$, is a quadratic function of the distance of the probe's tip from its target star when close to the goal, transitioning to a conic potential at distances greater than $d_\mathrm{T}$ from the target. The quadratic is desirable to ensure smooth deceleration (continuous first and second derivatives) near the goal, while the long-range behaviour of the conic ensures that the strength of the attraction does not increase without bound (the magnitude of the gradient is constant). Continuous first derivatives are also used as a boundary condition at the transition between the two regimes:
\begin{equation}
U_{i,\mathrm{att}}(\vec{q_i}) = \left\{
  \begin{array}{lll}
    \frac{1}{2} \beta d^2(\vec{q_i},\vec{q}_{i,\mathrm{goal}})&,& d \le d_\mathrm{T} \\
    d_\mathrm{T} \beta d(\vec{q_i},\vec{q}_{i,\mathrm{goal}}) -
      \frac{1}{2} \beta d_\mathrm{T}^2 &,& d > d_\mathrm{T}
  \end{array}
  \right.
\end{equation}
Here $d(\vec{q_i},\vec{q}_{i,\mathrm{goal}})$ represents the Cartesian distance between the current and goal locations for the probe tips, and $\beta$ is a weighting factor.

\begin{figure}[ht]
\begin{center}
\begin{tabular}{cc}
\includegraphics[width=0.49\linewidth]{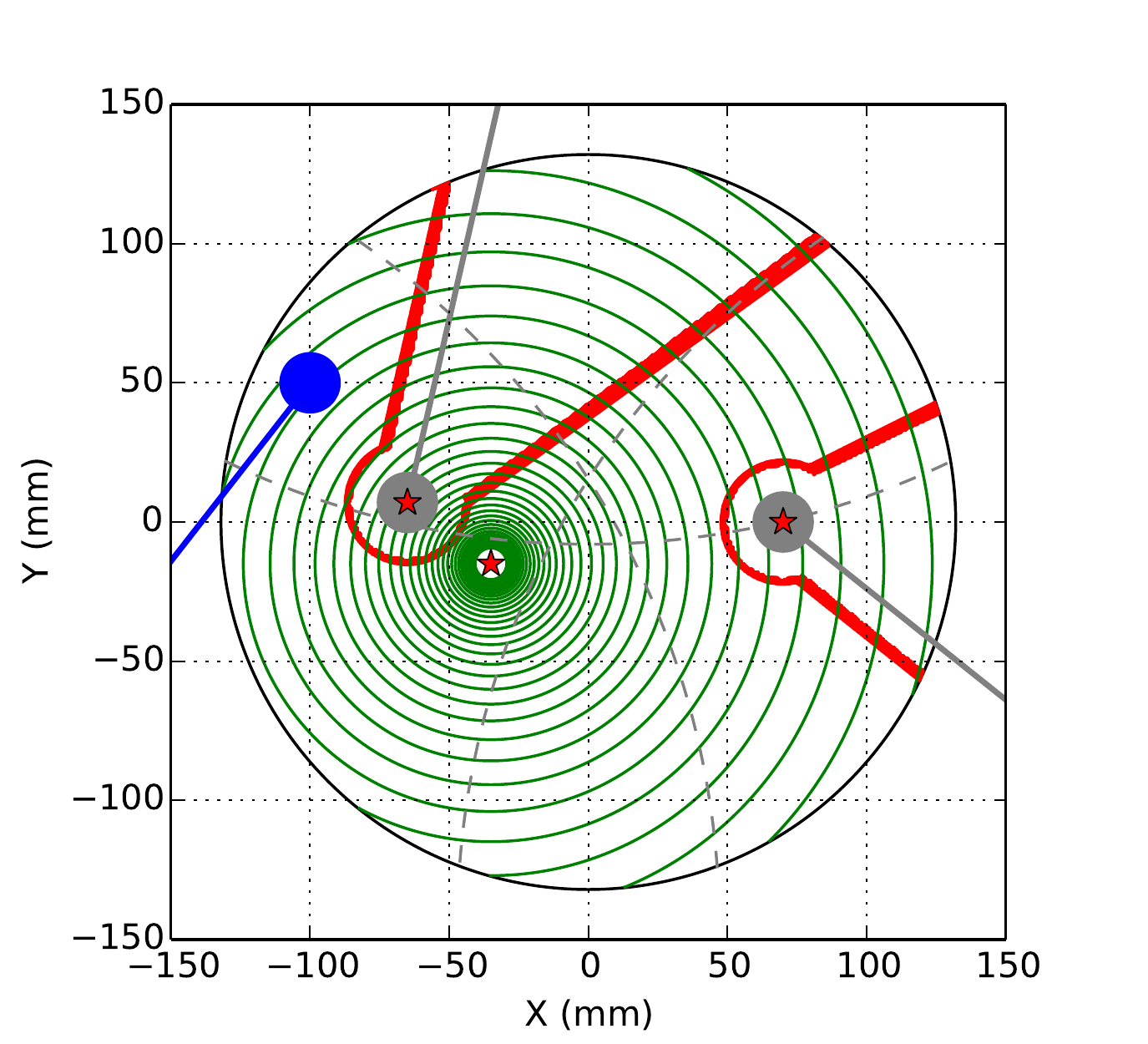} &
\includegraphics[width=0.49\linewidth]{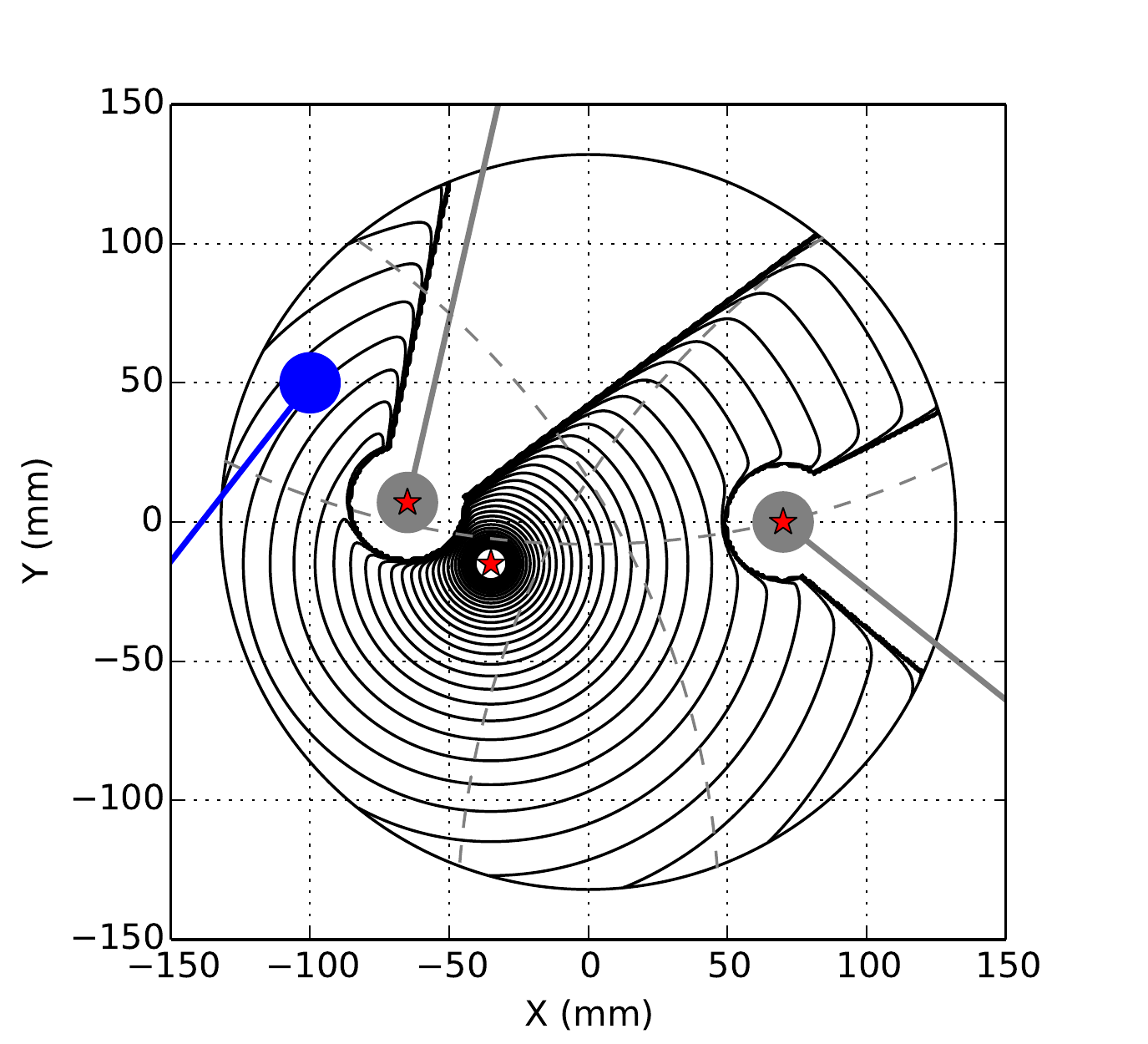}
\end{tabular}
\end{center}
\caption{\label{fig:potential}
\textit{left:} Depiction of the attractive (green contours) and repulsive potentials (red contours) experienced by the left-hand probe (blue). The remaining probes (grey) have already acquired their respective guide stars. The repulsive potential has a steep gradient, and is only short range, such that the contours blend into each other. \textit{right:} The combined potential, illustrating how the repulsive potential excludes regions of the patrol area that would result in collisions.}
\end{figure}

Finally, the negative gradient of the combined potential at the location of the $i$th probe's tip, $\vec{\psi}_{i,U}(\vec{q}_i)$, is used to identify the direction for the next step in its trajectory (i.e., new velocity demands):
\begin{equation}
  \vec{\psi}_{i,U}(\vec{q}_i) = -\nabla_{x,y} \left[
    \alpha U_{i,\mathrm{rep}}(\vec{q}_i) +
    \beta U_{i,\mathrm{att}}(\vec{q_i}) 
    \right] .
\end{equation}
In our simulations the potential and gradient are calculated in Cartesian $x,y$ coordinates (simplifying the calculations, and also making them easier to visualize). The gradient is then converted into polar-coordinate velocities in order to generate the next set of demands for each probe, $\vec{v} = (\dot{r}, \dot{\theta})$. This is achieved by projecting $\vec{\psi}_{i,U}$ along Cartesian unit vectors representing the probe's rotated coordinate system at its tip, $\hat{r}_i$ and $\hat{\theta}_i$:
\begin{equation}
       \hat{r}_i = (   \cos \theta_i, \sin \theta_i ),
  \hat{\theta}_i = ( - \sin \theta_i, \cos \theta_i ) ,
\end{equation}
where $\theta_i$ is the current probe rotation, and
\begin{equation}
  \vec{v}_i = (\vec{\psi}_{i,U} \cdot \hat{r}_i,
               \vec{\psi}_{i,U} \cdot \hat{\theta}_i/r_i ) .
\label{eq:demand}
\end{equation}
Fig.~\ref{fig:potential} shows the potential function (and its components) experienced by a single probe, in an example configuration for which the other two probes are stationary (already tracking their target stars). The adopted parameters for the potential are given in Table~\ref{tab:potential}. Note the short range of the repulsive potential.

\begin{table}[ht]
\caption{Motion planning algorithm parameters.}
\label{tab:potential}
\begin{center}
\begin{tabular}{|c|c|l|} 
\hline
Parameter & Description & Value \\
\hline
$\alpha$ & repulsive potential weight  & 10 \\
$Q^*$    & repulsive potential range   & 22\,mm \\
$d_\mathrm{T}$ & quadratic-conic transition distance & 5\,mm \\
$\beta$  & attractive potential weight & 0.001 \\
$\gamma$ & transverse component scale & 3 \\
\hline
\end{tabular}
\end{center}
\end{table}

\subsection{Mitigating local minima}

\begin{figure}[ht]
\begin{center}
\begin{tabular}{cc}
\includegraphics[width=0.49\linewidth]{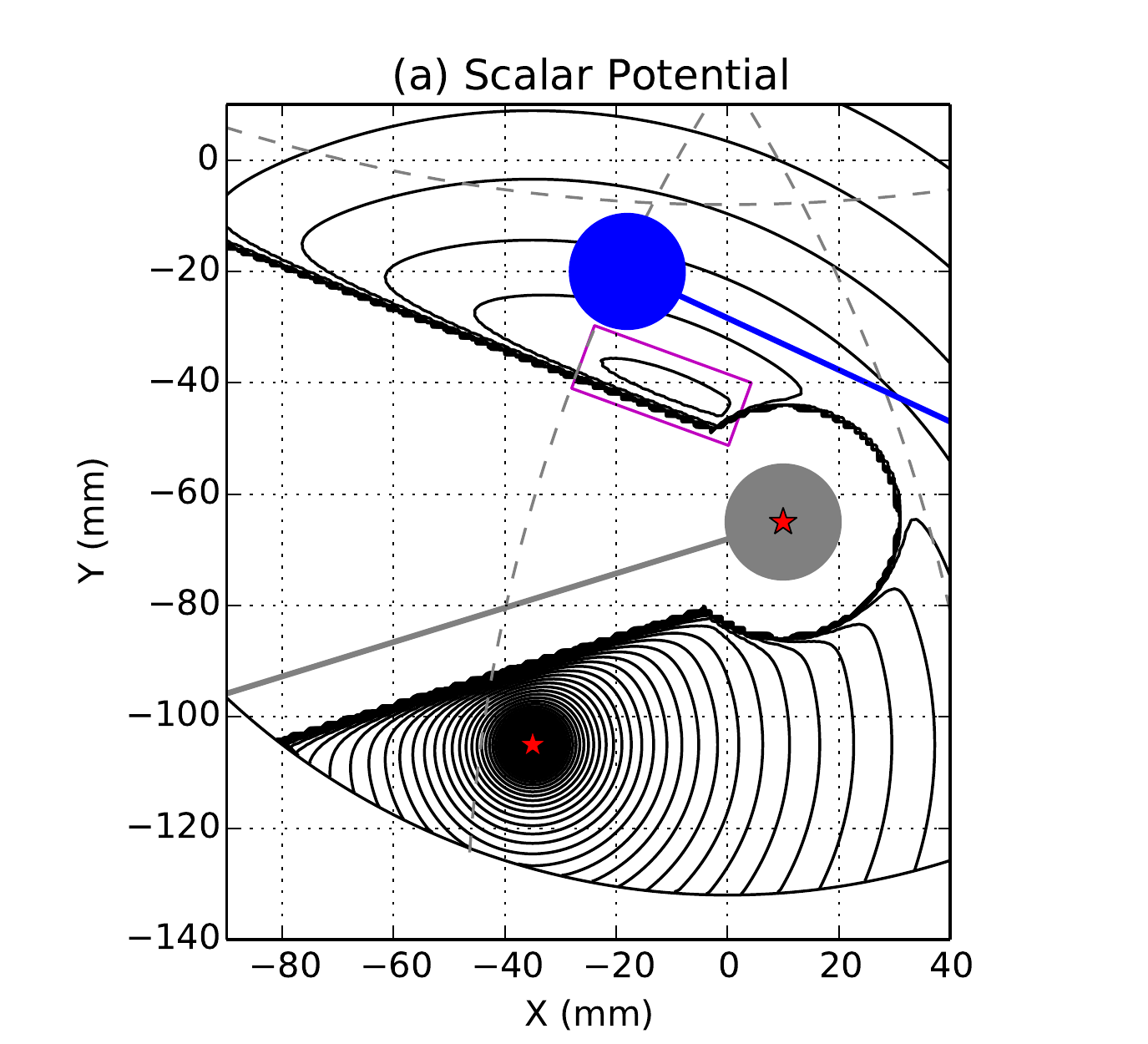} &
\includegraphics[width=0.49\linewidth]{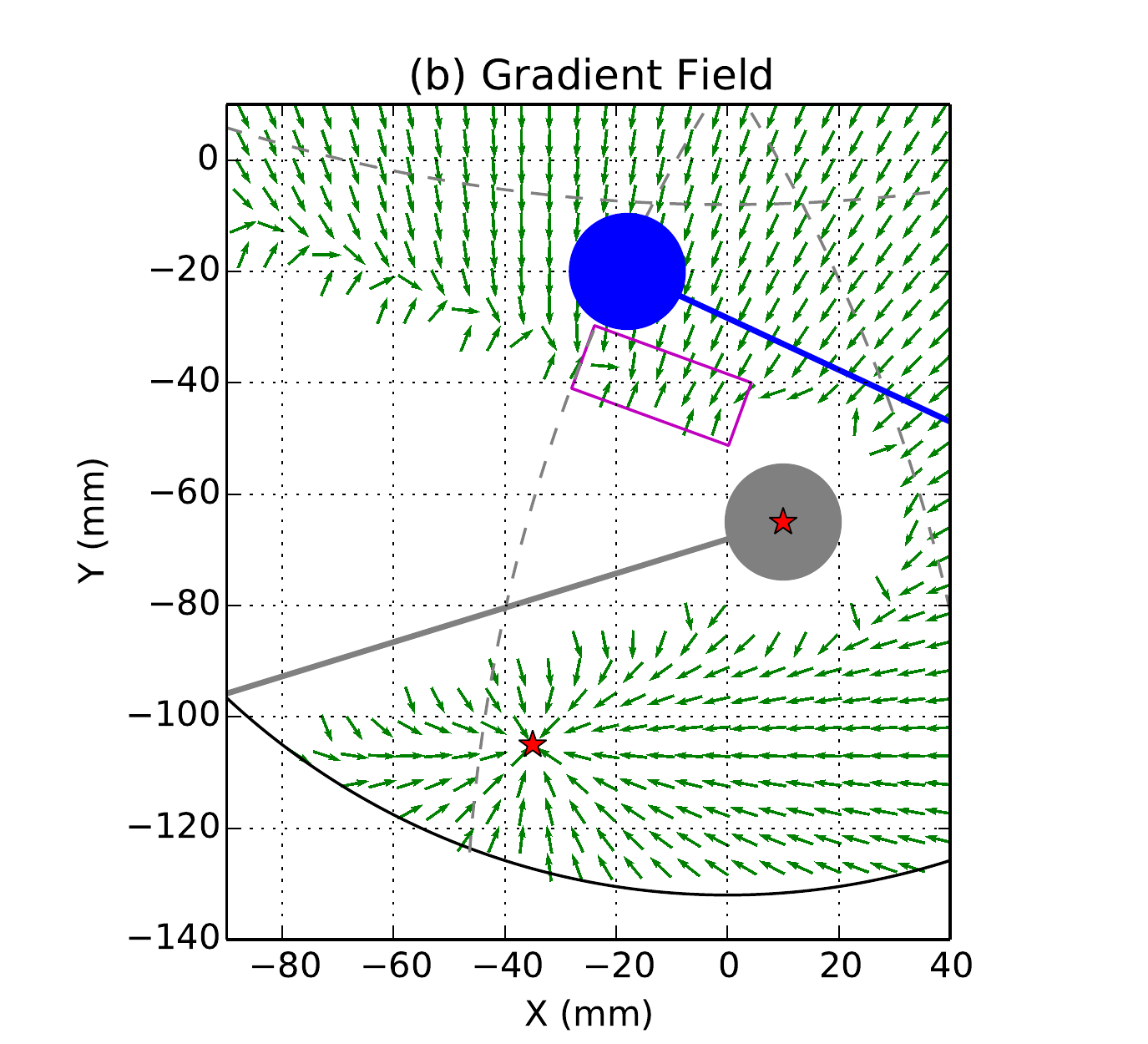} \\
\includegraphics[width=0.49\linewidth]{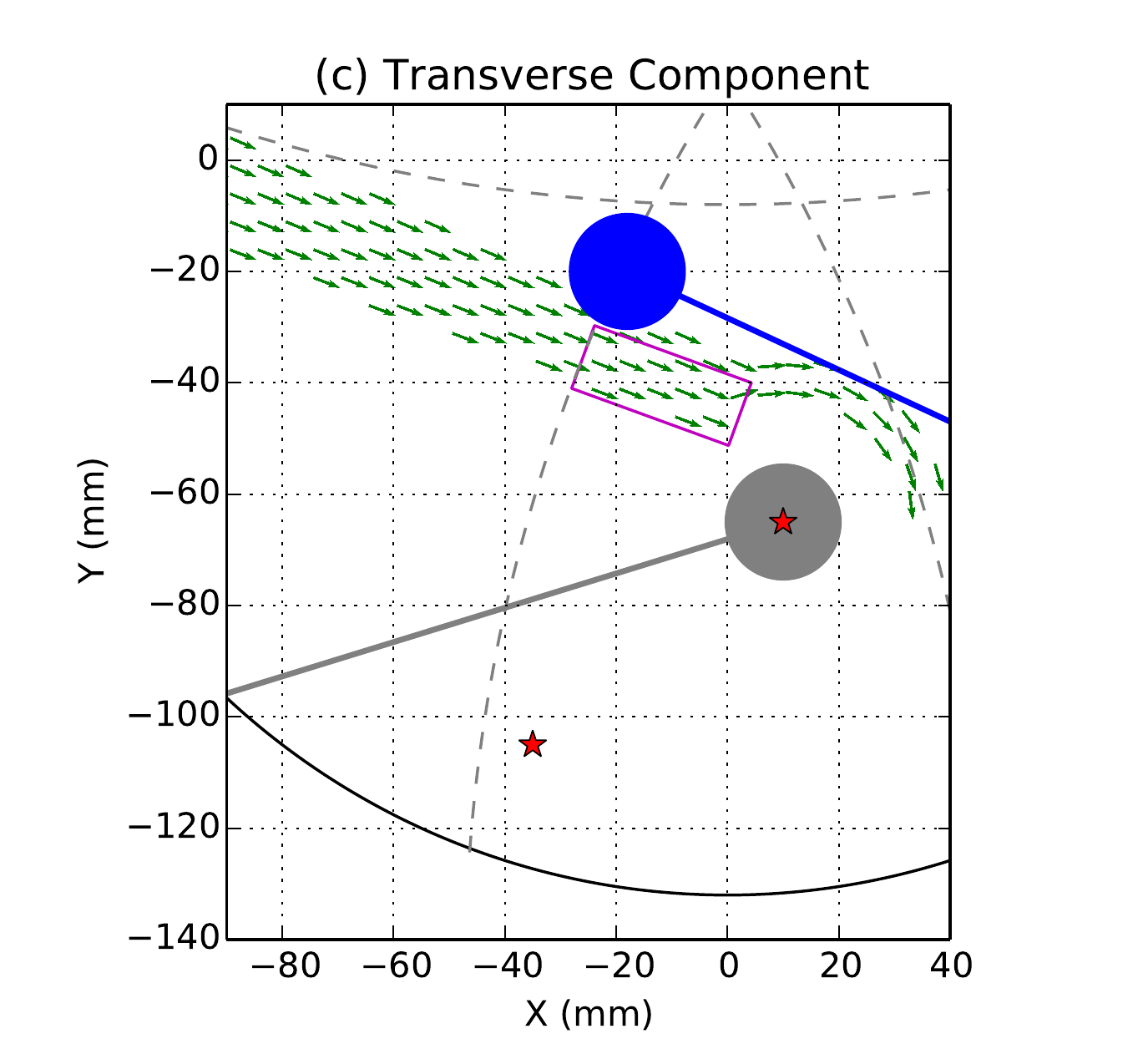} &
\includegraphics[width=0.49\linewidth]{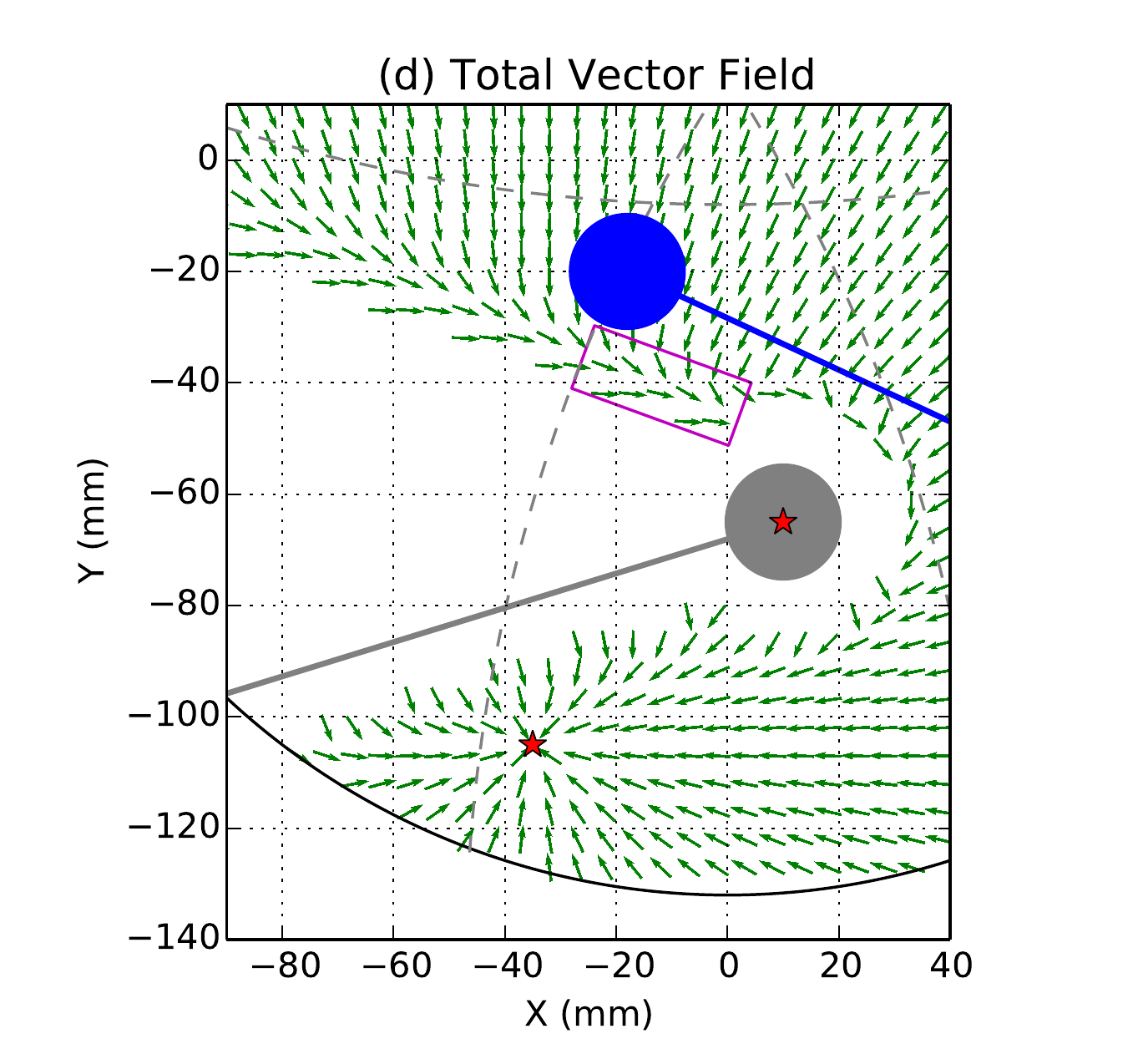} \\
\end{tabular}
\end{center}
\caption{\label{fig:vect}
(a) A scenario in which the combination of the attractive and repulsive potentials results in a local minimum (within the magenta rectangle) in the scalar potential (black contours) far from the goal position [near $(-35,-105)$], experienced by the top probe (blue). (b) Negative gradient of the potential (normalized to emphasize direction), illustrating the same feature (arrows point toward each other at the location of the local minimum). (c) Component added to the vector field that is transverse to, and proportional in magnitude to, the repulsive component, wherever the straight-line trajectory of a probe tip to its goal intersects another probe. (d) Complete vector field used to calculate motion, illustrating smooth streams past the problem area of panel (b).}
\end{figure}

While the algorithm presented above is conceptually simple, in practice, the combination of attractive and repulsive potentials can produce local minima far from the goal configuration, causing the probe to become \textit{deadlocked}. Indeed, for the OIWFS control problem, this is the case, as shown in panels (a) and (b) of Fig.~\ref{fig:vect}. These minima occur wherever the gradient of the attractive potential is the same magnitude as the gradient of the repulsive potential, but acting in precisely the opposite direction. This situation typically arises when one probe is extended significantly past the tip of another probe, and must move around it in order to reach the target star, as in the situation depicted.

The occurrence of local minima is highly dependent on the geometry of the problem. For example, a similar formulation to that used here was implemented to control arrays of ``$\theta-\phi$'' positioners for a fiber-fed spectrograph, but tuning the potential function in such a way that a single global minimum is guaranteed\cite{makarem2014a}. Such a potential is called a \textit{navigation function}. The theory of navigation functions is explored in detail in Ref.~\citenum{rimon1992}, for which the authors advocate a particular form of the potential in a ``sphere world'' configuration space (with spherical obstacles), and some possibilities for transforming it to serve other problem geometries. While methods such as these were considered for the OIWFS, we were unable to find a simple extension to our polar-coordinate robot problem. Instead, we have found a solution whereby an additional transverse component is added directly to the vector field.

\begin{figure}[ht]
\begin{center}
\includegraphics[width=0.8\linewidth]{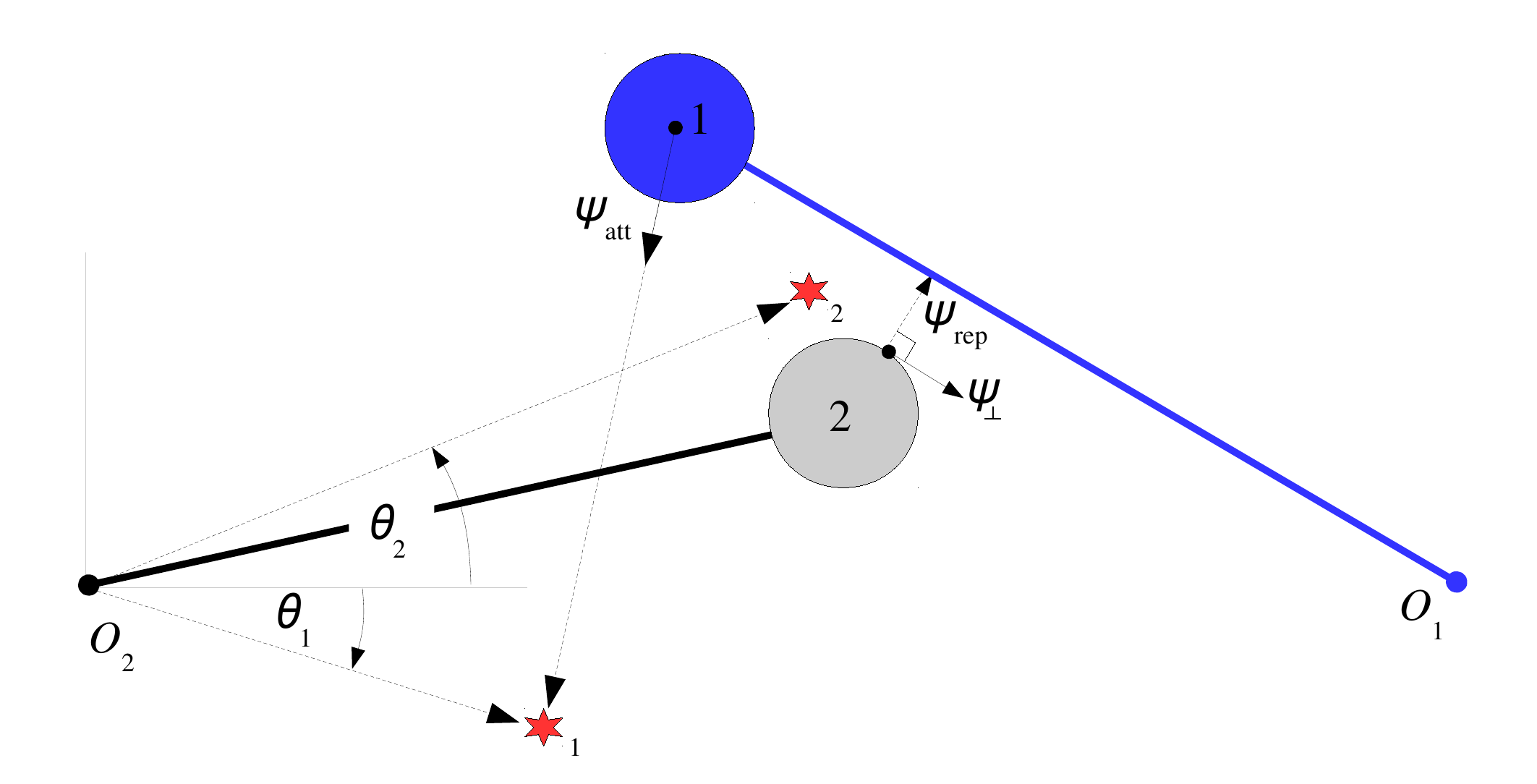}
\end{center}
\caption{\label{fig:tran}
Method for calculating the transverse component of the vector field, $\vec{\psi}_\perp$. If a straight line segment between the tip of Probe 1 and its target Star 1 intersects Probe 2, the magnitude of the transverse component is proportional to the downhill gradient of the repulsive potential, $\psi_\mathrm{rep}$, but applied orthogonally (it is zero otherwise). The sign is chosen based on the relative angles of the target stars for the two probes about Probe 2's origin (clockwise in this case). The velocity demand for Probe 1 is the vector sum $\vec{\psi}_\mathrm{att} + \vec{\psi}_\mathrm{rep} + \vec{\psi}_\perp$ applied to its head, but converted into polar coordinates.}
\end{figure}

Fig.~\ref{fig:tran} shows a situation schematically similar to that of Fig.~\ref{fig:vect}, although in this case, for illustration, neither of the probes have reached their targets. Considering the problem from the point of view of Probe 1, it is clear that it would collide with Probe 2 if it were to move directly towards its target. These situations are easy to detect algebraically; one must simply check for intersections between the straight line segment representing travel, with the line segments and circles representing other probes. Wherever these cases are encountered, a new component is added to the vector field, a simple scaling of the gradient of the repulsive potential, $\psi_\mathrm{rep}$ by a factor $\gamma$, but applied orthogonally:
\begin{equation}
  \vec{\psi}_{i,\mathrm{tot}}(\vec{q}_i) =
  \left[ \vec{\psi}_{i,U}(\vec{q}_i) + \vec{\psi}_{i,\perp}(\vec{q}_i)  \right] ,
\end{equation}
where
\begin{equation}
\vec{\psi}_{i,\perp}(\vec{q}_i) =
s \gamma \times \rot(\vec{\psi}_{i,\mathrm{rep}},90^{\circ}) ,
\end{equation}
and $s$, the sign, is either $+1$ or $-1$.

The choice of sign causes the transverse component to induce either clockwise ($s=-1$) or counter-clockwise ($s=+1$) motion of Probe 1 about Probe 2 (noting that the gradient of the repulsive potential, $\vec{\psi}_{1,\mathrm{rep}}$, always points away from Probe 2). A simple method to determine this sign geometrically is to compare the angles of both target stars when expressed in polar coordinates about the origin of Probe 2. If $\theta_1$ is less than $\theta_2$, then Star 1 lies at a clockwise rotational offset from Star 2; this is the same direction of travel required for Probe 1 (i.e., $s=-1$).

Note in Fig.~\ref{fig:tran} that $\vec{\psi}_{i,\mathrm{rep}}$, and $\vec{\psi}_{i,\perp}$ have been drawn such that they originate at the point on Probe 2 that is closest to Probe 1 for illustration: the direction of the repulsion experienced by Probe 1 from Probe 2 is parallel to the shortest line segment connecting them. In practice, these two components are summed with $\vec{\psi}_{i,\mathrm{att}}$ and applied to the tip of Probe 1.

Panels (c) and (d) of Fig.~\ref{fig:vect} demonstrate the effect this new component has on the vector field of velocity demands. The local minimum above the grey (stationary) probe is now replaced with a smooth stream that will carry the blue probe around it. In order to emphasize the direction of the vector field (instead of magnitude), the lengths of the arrows have been normalized. In practice, the magnitudes of the repulsive and transverse components smoothly decrease to zero a short distance from the grey probe.

\subsection{Simulations}

\begin{figure}[ht]
\begin{center}
\begin{tabular}{cc}
\includegraphics[width=0.49\linewidth]{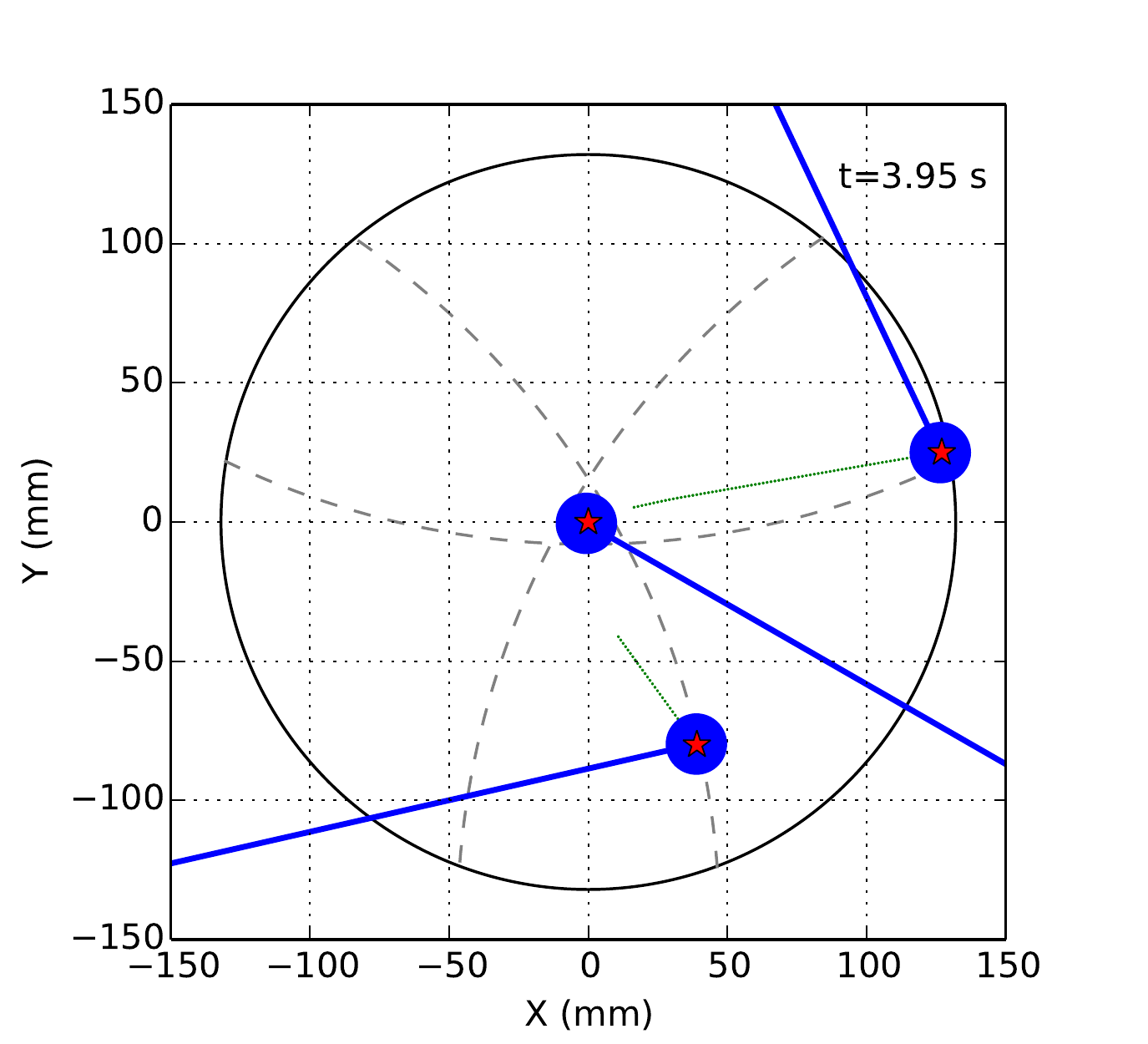} &
\includegraphics[width=0.49\linewidth]{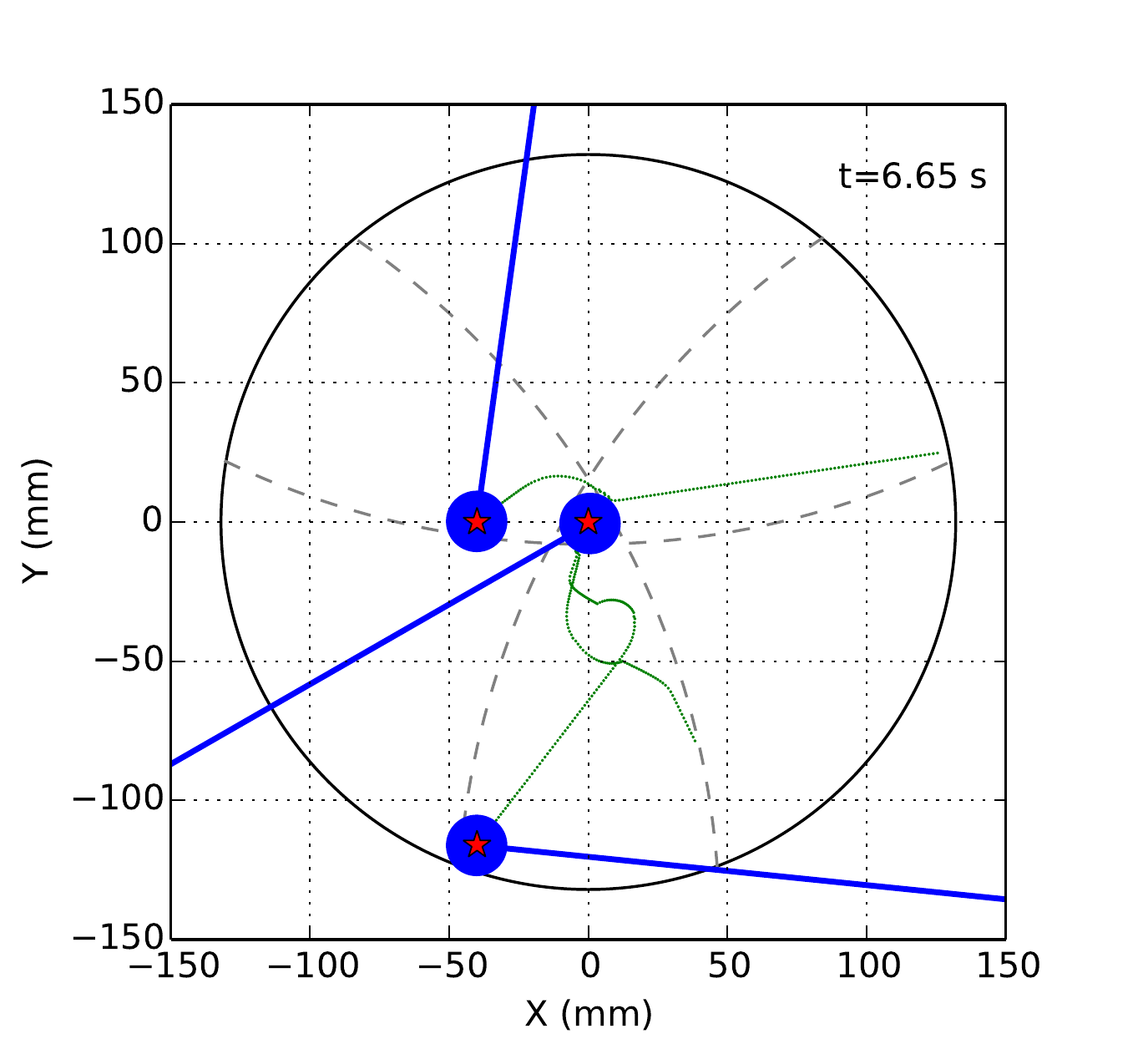} \\
\includegraphics[width=0.49\linewidth]{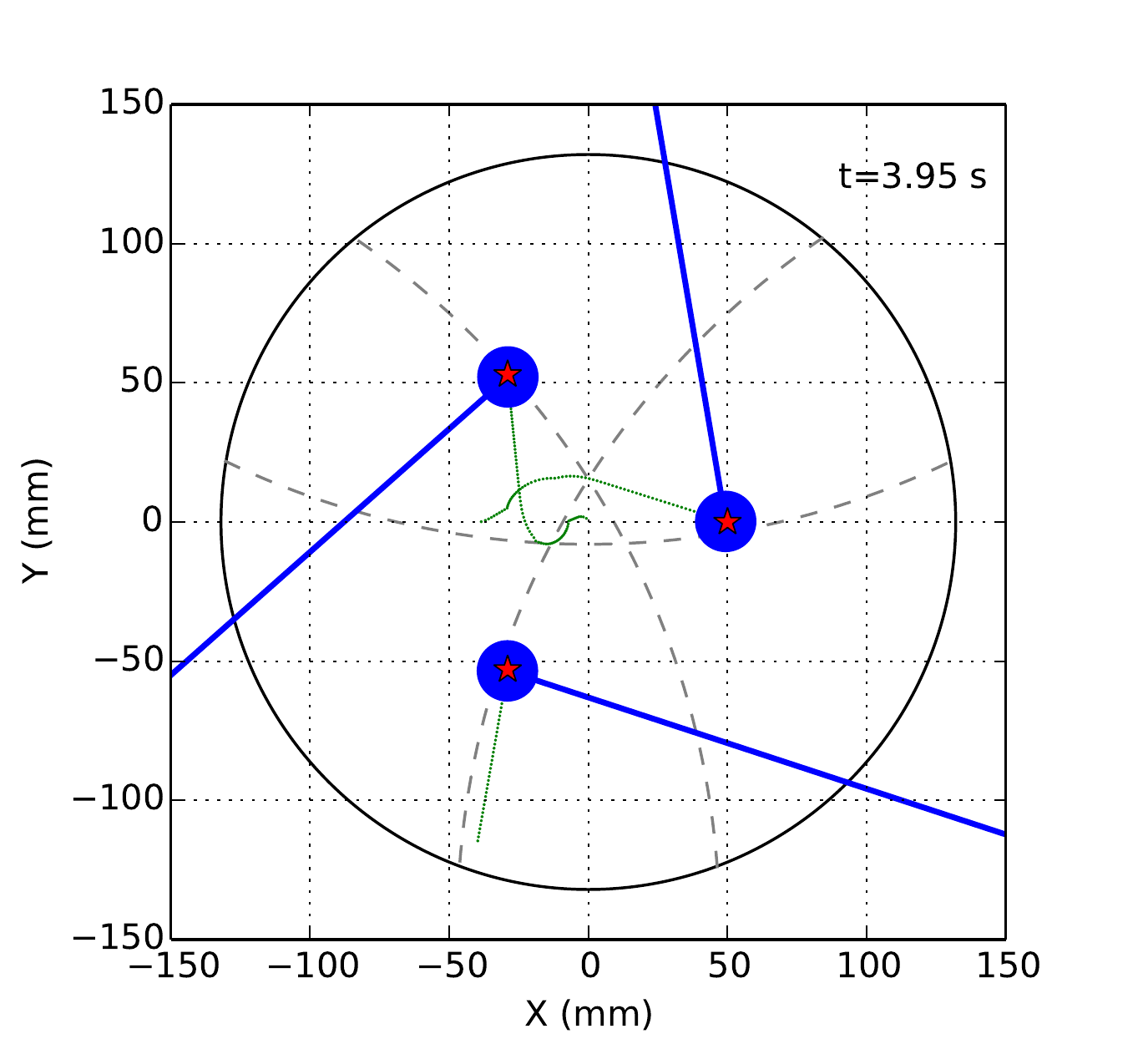} &
\includegraphics[width=0.49\linewidth]{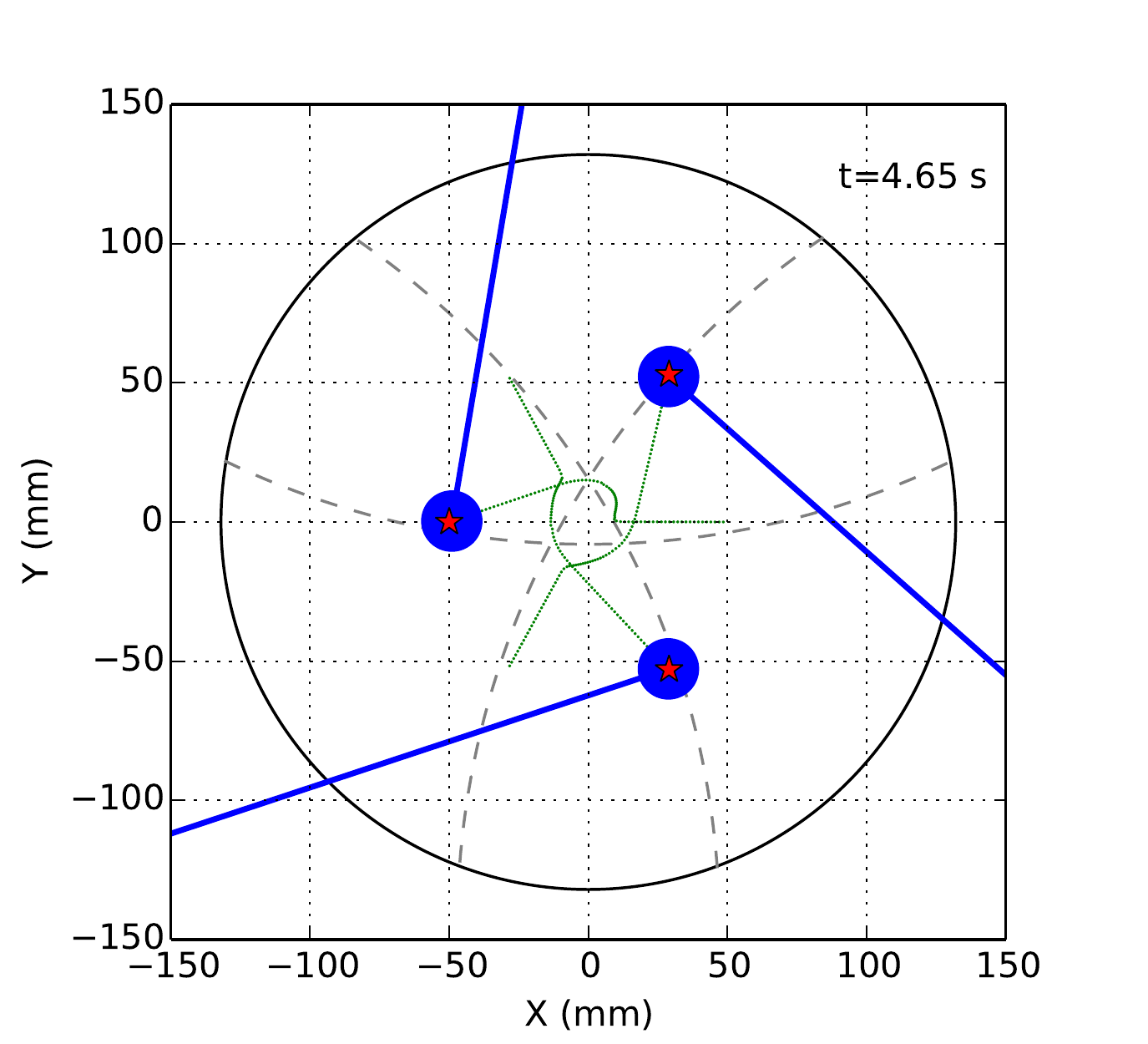} \\
\end{tabular}
\end{center}
\caption{\label{fig:reconfig}
A series of probe reconfigurations, from top-left to bottom-right. The numbers in the top right-hand corner of each panel indicate the time for each move. The green trails show the paths that each probe follows; deviations from straight lines demonstrate where manoeuvres were required to avoid collisions.}
\end{figure}

Next, we simulate the motion associated with a range of hand-selected probe reconfigurations using the modified vector field of velocity demands, as shown in Fig.~\ref{fig:reconfig}. We have made the simplifying assumption that the probe acceleration time is minimal. The maximum velocity of each motorized stage is chosen to meet the ``minimum tracking speed'' requirement in Tab~\ref{tab:model}: the linear stage is able to move at a maximum speed of 13.2\,mm/s, and the rotational stage is able to move at a maximum angular speed that produces 13.2\,mm/s of linear motion when the probe is at its minimum operational extension (at the edge of the FOV). Each vector velocity demand (Eq.~\ref{eq:demand}, but now using $\vec{\psi}_{i,\mathrm{tot}}$ in place of $\vec{\psi}_{i,U}$) is simply scaled to the largest value possible without exceeding the maximum in either axis at each step of the simulation. Note that, since both axes are capable of meeting the tracking speed requirement independently, when they are both in motion, the linear speed of the probe tip will generally exceed this requirement.

The initial panel (top-left) depicts a situation in which the probes are able to move directly to their destinations, as evidenced by the straight green trails. In the remaining panels, however, deviations in the trails indicate locations where manoeuvres were required to avoid collisions. The final panel (at the bottom-right) shows a particularly contrived probe reconfiguration designed to demonstrate the behaviour of the algorithm when avoiding a 3-way collision at the center of the field.

Finally, we generate a sequence of 1000 random reconfigurations of the probes. For this simulation it is important to note that there may be several valid sets of probe assignments for a given asterism, requiring a separate algorithm to decide amongst them. For example, one might simulate moves from the current probe configuration to each of the $N$ valid target configurations for the next asterism, and simply choose that which takes the shortest amount of time. Using such logic, cases like that depicted in Fig.~\ref{fig:tran} would often be avoided: in all likelihood, Probe 1 would be assigned to Star 2, and Probe 2 to Star 1 (i.e., Probe 1 would not have to move around Probe 2). However, in some complex operational scenarios, such reconfigurations may indeed be required. Consider, for example, a non-sidereal tracking case in which the guide stars of Fig.~\ref{fig:tran} are slowly moving upwards, and Probe 2 is tracking Star 2. If Probe 1 were tracking a star that had just drifted off the top of the FOV, the AO system may continue to operate in a closed-loop (though degraded) mode using only Star 2, while simultaneously moving Probe 1 to acquire the new Star 1 that has just drifted into the bottom of the field.

Ultimately, it will be the responsibility of higher-level observatory software (e.g., an observation planning tool) to make initial guesses at the asterisms and probe assignments, and to offer the observer the opportunity to change them. From the point of view of the OIWFS, it must simply respond to any valid position demands (ones that will not result in probe collisions at their goal locations). For this reason, our simulation of 1000 reconfigurations uses both random asterisms (selected uniformly across the FOV) and random probe assignments, to properly sample the full space of possibilities. We find that the median reconfiguration time is 6.25\,s (ranging from a minimum of 1.35\,s to a maximum of 10.4\,s).

\section{Summary and future work}

We have developed a simulation for IRIS OIWFS motion control using artificial potential functions (with attractive and repulsive components) for its Preliminary Design Phase. A key part of this simulation is the mitigation of local minima by adding to the vector field (that results from evaluating the negative gradient of the potential) a transverse component that acts orthogonally to the repulsive potential. This method is computationally inexpensive, and elegant in the sense that a single algorithm can be used both for initial configuration, and continuous motion in a closed-loop operational state, without special-case logic. Reconfiguration time statistics demonstrate that this algorithm should meet or exceed the design requirements for the OIWFS, and will help to keep its contribution to AO acquisition time overheads to a minimum. Future work will include modelling a realistic proportional-integral-derivative (PID) feedback loop for motor control, and optimizing the model parameters.

\acknowledgments

The TMT Project gratefully acknowledges the support of the TMT collaborating institutions.  They are the California Institute of Technology, the University of California, the National Astronomical Observatory of Japan, the National Astronomical Observatories of China and their consortium partners, the Department of Science and Technology of India and their supported institutes, and the National Research Council of Canada.  This work was supported as well by the Gordon and Betty Moore Foundation, the Canada Foundation for Innovation, the Ontario Ministry of Research and Innovation, the Natural Sciences and Engineering Research Council of Canada, the British Columbia Knowledge Development Fund, the Association of Canadian Universities for Research in Astronomy (ACURA) , the Association of Universities for Research in Astronomy (AURA), the U.S. National Science Foundation, the National Institutes of Natural Sciences of Japan, and the Department of Atomic Energy of India. EC would also like to thank Dan Kerley for his comments on the manuscript.

\bibliography{refs} 

\begin{thebibliography}{1}

\bibitem{larkin2016}
{Larkin}, J.~E., {Moore}, A.~M., {Wright}, S.~A., {Wincentsen}, J.~E.,
  {Chisholm}, E.~M., {Dekany}, R.~G., {Dunn}, J.~S., {Ellerbroek}, B.~L.,
  {Hayano}, Y., {Phillips}, A.~C., {Simard}, L., {Smith}, R.~M., {Suzuki}, R.,
  {Weiss}, J.~L., and {Zhang}, K., ``{The Infrared Imaging Spectrograph (IRIS)
  for TMT: Instrument Overview},'' in [{\em Ground-based and Airborne
  Instrumentation for Astronomy VI}{\nolinebreak\hspace{0.1em}]},  {\em \spie}
  {\bf 9908},  990870 (July 2016).

\bibitem{herriot2014}
{Herriot}, G., {Andersen}, D., {Atwood}, J., {Boyer}, C., {Byrnes}, P.,
  {Caputa}, K., {Ellerbroek}, B., {Gilles}, L., {Hill}, A., {Ljusic}, Z.,
  {Pazder}, J., {Rosensteiner}, M., {Smith}, M., {Spano}, P., {Szeto}, K.,
  {V{\'e}ran}, J.-P., {Wevers}, I., {Wang}, L., and {Wooff}, R., ``{NFIRAOS:
  first facility AO system for the Thirty Meter Telescope},'' in [{\em Adaptive
  Optics Systems IV}{\nolinebreak\hspace{0.1em}]},  {\em \spie} {\bf 9148},
  914810 (July 2014).

\bibitem{dunn2016}
{Dunn}, J., {Andersen}, D., {Chapin}, E., {Reshetov}, V., {Wierzbicki}, R.,
  {Herriot}, G., {Chalmers}, D., {Isbrucker}, V., {Larkin}, J.~E., {Moore},
  A.~M., and {Suzuki}, R., ``{The Infrared Imaging Spectrograph (IRIS) for TMT:
  Multi-tiered Wavefront Measurements and Novel Mechanical Design},'' in [{\em
  Ground-based and Airborne Instrumentation for Astronomy
  VI}{\nolinebreak\hspace{0.1em}]},  {\em \spie} {\bf 9908},  9908381 (July
  2016).

\bibitem{goodwin2014}
{Goodwin}, M., {Lorente}, N.~P.~F., {Satorre}, C., {Hong}, S.~E., {Kuehn}, K.,
  and {Lawrence}, J.~S., ``{Field target allocation and routing algorithms for
  Starbugs},'' in [{\em Software and Cyberinfrastructure for Astronomy
  III}{\nolinebreak\hspace{0.1em}]},  {\em \spie} {\bf 9152},  91520S (July
  2014).

\bibitem{sugai2015}
Sugai, H., Tamura, N., Karoji, H., Shimono, A., Takato, N., Kimura, M., Ohyama,
  Y., Ueda, A., Aghazarian, H., de~Arruda, M.~V., Barkhouser, R.~H., Bennett,
  C.~L., Bickerton, S., Bozier, A., Braun, D.~F., Bui, K., Capocasale, C.~M.,
  Carr, M.~A., Castilho, B., Chang, Y.-C., Chen, H.-Y., Chou, R. C.~Y., Dawson,
  O.~R., Dekany, R.~G., Ek, E.~M., Ellis, R.~S., English, R.~J., Ferrand, D.,
  Ferreira, D., Fisher, C.~D., Golebiowski, M., Gunn, J.~E., Hart, M., Heckman,
  T.~M., Ho, P. T.~P., Hope, S., Hovland, L.~E., Hsu, S.-F., Hu, Y.-S., Huang,
  P.~J., Jaquet, M., Karr, J.~E., Kempenaar, J.~G., King, M.~E., Fèvre, O.~L.,
  Mignant, D.~L., Ling, H.-H., Loomis, C., Lupton, R.~H., Madec, F., Mao, P.,
  Marrara, L.~S., Ménard, B., Morantz, C., Murayama, H., Murray, G.~J.,
  de~Oliveira, A.~C., de~Oliveira, C.~M., de~Oliveira, L.~S., Orndorff, J.~D.,
  de~Paiva~Vilaça, R., Partos, E.~J., Pascal, S., Pegot-Ogier, T., Reiley,
  D.~J., Riddle, R., Santos, L., dos Santos, J.~B., Schwochert, M.~A.,
  Seiffert, M.~D., Smee, S.~A., Smith, R.~M., Steinkraus, R.~E., Sodré, Jr.,
  L., Spergel, D.~N., Surace, C., Tresse, L., Vidal, C., Vives, S., Wang,
  S.-Y., Wen, C.-Y., Wu, A.~C., Wyse, R., and Yan, C.-H., ``Prime focus
  spectrograph for the subaru telescope: massively multiplexed optical and
  near-infrared fiber spectrograph,'' {\em Journal of Astronomical Telescopes,
  Instruments, and Systems}~{\bf 1}(3),  035001 (2015).

\bibitem{makarem2014a}
{Makarem}, L., {Kneib}, J.-P., {Gillet}, D., {Bleuler}, H., {Bouri}, M.,
  {Jenni}, L., {Prada}, F., and {Sanchez}, J., ``{Collision avoidance in
  next-generation fiber positioner robotic systems for large survey
  spectrographs},'' {\em \aap}~{\bf 566},  A84 (June 2014).

\bibitem{latombe1991}
Latombe, J.-C.,  [{\em Robot Motion Planning}{\nolinebreak\hspace{0.1em}]},
  Kluwer Academic Publishers, Norwell, MA, USA (1991).

\bibitem{rimon1992}
Rimon, E. and Koditschek, D.~E., ``Exact robot navigation using artificial
  potential functions,'' {\em IEEE Transactions on Robotics and
  Automation}~{\bf 8},  501--518 (Oct 1992).

\end{thebibliography}
\bibliographystyle{spiebib} 

\end{document}